\documentclass[pre,aps,
notitlepage,
pdflatex,
superscriptaddress,floats,floatfix,showpacs,10pt]{revtex4-1}
\usepackage{graphicx}
\usepackage{amssymb}
\usepackage{amsmath}
\usepackage{textcomp}
\usepackage{color}

\newcounter{firstbib}

\begin{document}

\title{Radiation Dominated Electromagnetic Shield}
\author{S. V. Bulanov$^{1}$, T. Zh. Esirkepov$^{1}$, S. S. Bulanov$^{2}$, J. K. Koga$^1$, 
 K. Kondo$^1$, 
and M. Kando$^1$\\
$^1${\small {Kansai Photon Science Institute, 
National Institutes for Quantum and Radiological Science and Technology (QST), 
8-1-7 Umemidai, Kizugawa, Kyoto 619-0215, Japan}}\\
$^2${\small {Lawrence Berkeley National Laboratory, Berkeley, California 94720, USA}}\\
}

\begin{abstract}
We analyze the collision of a high energy electron beam with 
an oscillating electric and magnetic field configuration, 
which represents a three-dimensional standing electromagnetic wave. 
The radiating electrons are stopped at the distance 
of the order of or less than the electromagnetic wave wavelength,
and become trapped near the electric field local maxima
due to the nonlinear dependence 
of the radiation friction force on the electromagnetic field strength,
while the quantum effects on the radiation friction remain negligible.
\end{abstract}

\pacs{52.38.-r, 41.60.-m, 52.27.Ep}
\maketitle

The multiple colliding laser pulses (MCLP) concept
formulated in Ref. \cite{SSB-2010a} 
has been considered
for achieving high intensity electromagnetic (EM) field regimes  
(see Refs. \cite{SSB-2010b, 
Gonoskov-2012, 
Gonoskov-2013, Vranic-2017, Gong-2017}). 
In this concept,
a laser beam is split into $N$ equal sub-beams,
which then combine in a constructive way.
The laser beam energy, $\mathcal{E}_1$, 
is related to its electric field, $E_1$, and intensity, $I_1$,
as $\mathcal{E}_1 \propto E_1^2 \propto I_1$
\cite{Bassett-1986}.
Each sub-beam receives the energy of $\mathcal{E}_1/N$,
and, with the same focusability,
it has
the electric field of $E_1/\sqrt{N}$
and the intensity of $I_1/N$.
A constructive combination of sub-beams
gives 
the electric field of $E_N=\sqrt{N} E_{1}$, 
and the intensity of $I_{N}=N I_{1}$,
which can be substantially higher than for a single unsplit beam.
For a large number of sub-beams,
the electric field at the focus region, $E_N$,
is obviously constrained by the diffraction limit.

In the near future, the MCLP concept realization with next generation lasers will enable experimental 
studies of novel physics, characterized by the significant role of the radiation friction, 
coming into play at substantially high electromagnetic radiation intensity. 
In radiation friction dominated regimes the charged particle dynamics becomes principally different 
from that in the relatively moderate  intensity limit \cite{MTB},
 leading, in particular,  to the
 generation of high power gamma-flashes during laser irradiation of plasma targets \cite{Gamma_R-2012}.
For potential applications of laser based gamma-ray sources see the review article \cite{GALES-2016}. 
Theoretical studies
 \cite{Gonoskov_2016}, show that the MCLP 
concept can also be beneficial for realizing such important laser-matter interaction regimes as,
 for example, electron-positron pair production 
via the Breit-Wheeler process \cite{Vranic-2017, Gong-2017}. 
In extreme intensity limits,  the  radiation friction effects on ion acceleration, magnetic field self-generation,  
high-order-harmonics, and electron self-injection have a significant impact
\cite{Tamburini}.

This paper presents results of the theoretical analysis of the radiating electron motion in 
a 3D electromagnetic field configuration  \cite{SSB-2010b},
corresponding to a large number  of colliding electromagnetic beams, $N\to \infty$, 
in the MCLP concept.
We consider TE-TM polarization, which is formed by a superposition  of the 
TE-mode with Toroidal Electric and poloidal magnetic fields 
and the TM-mode with Toroidal Magnetic and poloidal electric fields.
Particular attention is paid to the case when the electron beam 
is trapped in the regions of high electromagnetic field amplitude,
while the quantum effects on the particle dynamics are negligible.

In the above mentioned configurations,
the electric field maximum $E_N$ is proportional to the square root 
of the electromagnetic wave power $\mathcal{P}$ \cite{Bassett-1986}.
In terms of the normalized field amplitude,
$a_N=e E_N/m_e \omega c$,
this relationship becomes
$a_N = \sqrt{\mathcal{P}/\mathcal{P}_c}$.
Here $e$, $m_e$, and $c$ are the electron charge, mass and 
speed of light in vacuum, respectively.
The characteristic power $\mathcal{P}_c=3 m_e^2 c^5/32 e^2$ 
is approximately equal to 1.59 GW.
For 10 PW  radiation power, the normalized electromagnetic 
field amplitude maximum is about $2.5\times 10^{3}$. 

The toroidal magnetic and electric fields
in cylindrical coordinates $(\rho, \phi, z)$
can be expressed as
\begin{equation}
\left(\!\!
\begin{array}{c}
B_{\phi}  \\  E_{\phi}
\end{array}
\!\!\right)
=
\left(\!\!
\begin{array}{c}
a_{TM} \sin(t)  \\  a_{TE} \sin(t+\varphi_{TE})
\end{array}
\!\!\right)
\!
\sqrt{\frac{2}{\pi}}\rho
 \left [
\frac{\sin{R}-R\cos{R}}{R^{3}}
\right].
\label{eq:BE-phi-fields}
\end{equation}
where $R=\sqrt{\rho^2+z^2}$.
Here and below we measure the electromagnetic field in the units of $m_e \omega c/e$, $a_{TM}$ and $a_{TE}$ 
are the normalized amplitudes of the TM and TE modes, and $\varphi_{TE}$ is 
the phase difference between them.
We assume azimuthal symmetry, i.e. $\partial_{\phi}=0$. The variables $t$ and $R$ are normalized by 
$\omega^{-1}$ and $k=c/\omega$, respectively. 
In terms of Fourier components, the poloidal electric field is
${\bf E}=i k (\nabla \times {\bf B})$ in the case of the TM mode and
the poloidal magnetic field is  ${\bf B}=-i k (\nabla \times {\bf E})$ for the TE mode.
The maximum of the TM (TE) field is
$a_{TM }\sqrt{8/9\pi}\approx 0.531 a_{TM}$ ($a_{TE }\sqrt{8/9\pi}\approx 0.531 a_{TE}$).

The relativistic electron dynamics in the electromagnetic field is described by the equations of motion: 
\begin{equation}
\dot {\bf p}={\bf E}+ \dot {\bf x} \times {\bf B}+ {\bf g}_{rad},
\label{eq:equmot-mom}
\end{equation}
with $ \dot {\bf x}={\bf p}/\gamma $, where ``dot'' stands for differentiation with respect to time. 
Here ${\bf p}$ is the electron momentum measured in the units $m_ec$,  
and 
$\gamma=(1+{\bf p}^2)^{1/2}$ 
is the electron gamma-factor.
The radiation friction force, ${\bf g}_{rad}$,
taken in the Landau-Lifshitz form \cite{LL-TF},
can be written
in the large energy limit $\gamma \gg 1$ 
as ${\bf g}_{rad}=-(2 \, \alpha\, a_S \,  \chi_e^2)/(3\gamma){\bf p}$. 
Here $\alpha=e^2/\hbar c \approx 1/137$ is the fine structure constant, $a_S=1/k \lambdabar_C=m_ec^2/\hbar \omega=eE_S/m_e \omega c$, 
where $\lambdabar_C=\hbar/m_ec \approx 3.86\times 10^{-11}$cm, is the reduced Compton wavelength,  $\hbar$ is the reduced Planck constant, 
$E_S=m_e^2c^3/e\hbar$ is the critical QED electric field (or Sauter-Schwinger field). For $1\mu$m wavelength 
laser radiation the dimensionless parameter $a_S$ is of the 
order of $5.1\times10^5$. The QED parameter $\chi_e$ characterizing multi-photon Compton processes is defined as \cite{BLP-QED}: 
$\chi_e=\sqrt{|F^{\mu \nu} p_{\nu}|^2}/a_S$, where $p_\nu$  denotes  the 4-momentum 
of an electron normalized by $m_e c$, which is given by $p_{\nu}=(\gamma, {\bf p})$ , 
the 4-tensor of the electromagnetic field is defined as $F_{\mu \nu}=\partial_{\mu} A_{\nu}-\partial_{\nu} A_{\mu}$.  

The radiation friction force can be rewritten as:
\begin{equation}
\label{eq:fradthree dimension}
{\bf g}_{rad}=-\varepsilon_{rad}\,{\bf p}\,\gamma \left[\left({\bf E}+\,{\dot {\bf x}}\times {\bf B}\right)^2
	 -\left(\dot {\bf x}\cdot {\bf E}\right)^2\right].
\end{equation}
On the r.h.s. of this equation, the dimensionless parameter 
$\varepsilon_{rad}$ is defined as $\varepsilon_{rad}=2 k r_e/3 =2\alpha/3a_S$, where $r_e=e^2/m_e c^2 \approx2.82\times 10^{-13}$cm is the classical electron radius. 
For $1\mu$m wavelength laser radiation this is of the order of $1.2\times10^{-8}$.
We assume that the quantum effects on the radiation friction 
are negligibly weak, which implies  smallness  of the QED parameter  $\chi_e$.

The parameter $\varepsilon_{rad}$ characterizes the role of the radiation friction force 
on the dynamics of a radiating electron. For example, when an ultrarelativistic electron rotates in the anti-nodes 
of a circularly polarized electromagnetic wave, the power emitted is 
proportional to the fourth power of its energy \cite{LL-TF}: 
$ {\frak p}_{\gamma}=\varepsilon_{rad}\gamma^2(\gamma^2-1)$. Here ${\frak p}_{\gamma}$ 
is the electromagnetic radiation power normalized by ${\frak p}_{c,\gamma}=m_ec^2\omega$. 
The electron can acquire energy from the electromagnetic field with the rate (normalized on ${\frak p}_{c,\gamma}$) $\approx a$, 
i. e. its energy is approximately equal to $\gamma=a$, where $a$ is the normalized electromagnetic field amplitude. 
The condition of the balance between the acquired and lost energy yields $a^3\approx \varepsilon_{rad}^{-1}$. 
The radiation friction effects become dominant in the large EM field amplitude limit, when $a>a_{rad}=\varepsilon_{rad}^{-1/3}$. 
For $a\gg a_{rad}$ the electron energy scales as ${\cal E}_{rad}/m_ec^2 \approx (a/\varepsilon_{rad})^{1/4}$. For details see Ref. \cite{LAD-LL}  and references therein.

When an ultrarelativistic electron with initial momentum ${\bf p}_0$ crosses the region of a strong electromagnetic field it loses energy.
As easily obtained from Eq. (\ref{eq:equmot-mom}) the rate of energy loss is 
\begin{equation}
\dot {\gamma}={\dot {\bf x}}\cdot \left({\bf E}+ {\bf g}_{rad}\right).
\label{eq:equradgamma}
\end{equation}
Assuming that $p \gg a$, $a\gg 1$, i. e. we can represent the particle momentum as ${\bf p}=p\, {\bf e}_p $ 
with ${\bf e}_p$ being the unit vector in the direction of the momentum ${\bf p}_0$, and $|{\bf g}_{rad}|\gg a$, 
we can rewrite Eq. (\ref{eq:equradgamma}) as the equation for the electron momentum (see \cite{LL-TF, P40}) 
$\dot p =-\varepsilon_{rad}\,p^2\,\left[\left({\bf E}+{\bf e}_p\times {\bf B}\right)^2-\left({\bf e}_p\cdot {\bf E}\right)^2\right]$. 
As an example, considering the electron moving along the $z$ direction we can find 
\begin{equation}
p'=-\varepsilon_{rad}\,p^2\,\Pi_z(\rho,z)
\label{eq:radatZaxis}
\end{equation}
with ``prime'' denoting the differentiation with respect to the coordinate $z$.
Here it is taken into account that for an ultrarelativistic particle moving parallel to the $z$ axis $t=z$. 
The function $\Pi_z(\rho,z)=\left(E_{\rho}(\rho,z)-B_{\phi}(\rho,z)\right)^2+\left(E_{\phi}(\rho,z)+B_{\rho}(\rho,z)\right)^2 $ 
describes the space distribution of the radiation friction force acting on the ultrarelativistic electron beam propagating 
in the $z$ direction. 

In the case of the electron moving parallel to  the $z$ axis, if its trajectory is located a small distance $\rho_0\ll 1$ from the $z$-axis,  
integration of  Eq. (\ref{eq:radatZaxis})
yields
\begin{equation}
p_{\infty}=\frac{p_0}{1+\displaystyle (2/ \pi) p_0  \varepsilon_{rad}\, a_{0}^2 \rho_0^2 \Phi(\infty)},
\label{eq:radataxispINF}
\end{equation}
where $\Phi(\infty) = 16\pi/105$ (see Appendix).

In the high electromagnetic field and/or the high initial energy limit, the particle loses almost all its energy before entering the 
maximum field amplitude  region. Asymptotically, for $a_0^{2} p_0 \rho_0^2 \varepsilon_{rad} \gg 32/105$ the resulting 
particle energy, $\gamma_f \approx p|_{p_0\to\infty}$, is independent of its initial energy,  $\gamma_0 \approx p_0$, being equal to  
\begin{equation}
\gamma_f=\frac{105}{32 a_0^{2} \rho_0^2 \varepsilon_{rad}},
\label{eq:gammaf}
\end{equation}
in accordance with \cite{P40} and \cite{LL-TF} (see also Refs. \cite{
KEB-2005, design, Vranic-2014}).

\begin{figure}[tb]
    \includegraphics[width=0.66\columnwidth]{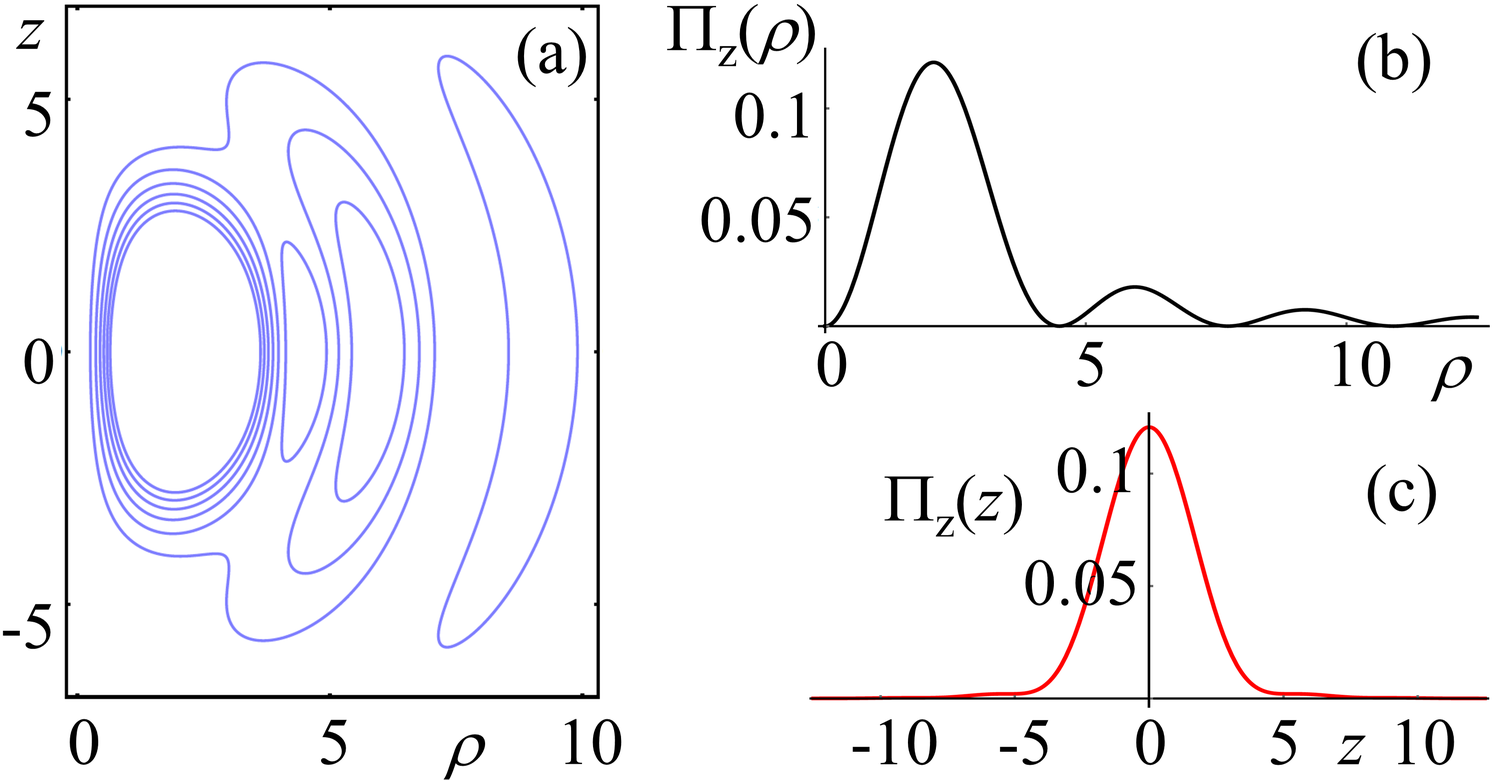}
    \caption{a) Isocontours of the function $\Pi_z$ in the $\rho,z$ plane.
b) $ \Pi_z(\rho)$ at $z=0$.
c) $ \Pi_z(z)$ at $\rho=2$.
}
\label{fig:1}
\end{figure}

We assume here and below for sake of brevity that $a_{TE}=a_{TM}=a_0$ and $\varphi_{TE}=0$. 
Using this assumption, in Fig. \ref{fig:1}, we plot the isocontours of the function $\Pi_z (\rho,z)$ in the $\rho,z$ plane (Panel (a)), 
 its  radial dependence at $z=0$ (Panel (b)), and its dependence on the coordinate $z$ for $\rho=2$ (Panel (c)).
The function $\Pi_z (\rho,z)$ vanishes along the axis $z$ for $\rho=0$, as seen in Figs. \ref{fig:1} (a) and (b). 
It is proportional to the square of the coordinate $\rho$ for  
$\rho \ll 1$. It reaches a maximum at $\rho$ approximately equal to 2, where $\Pi_z \approx 0.12$ 
(see Figs. \ref{fig:1} (b) and (c)).

\begin{figure}[tb]
    \includegraphics[width=0.66\columnwidth]{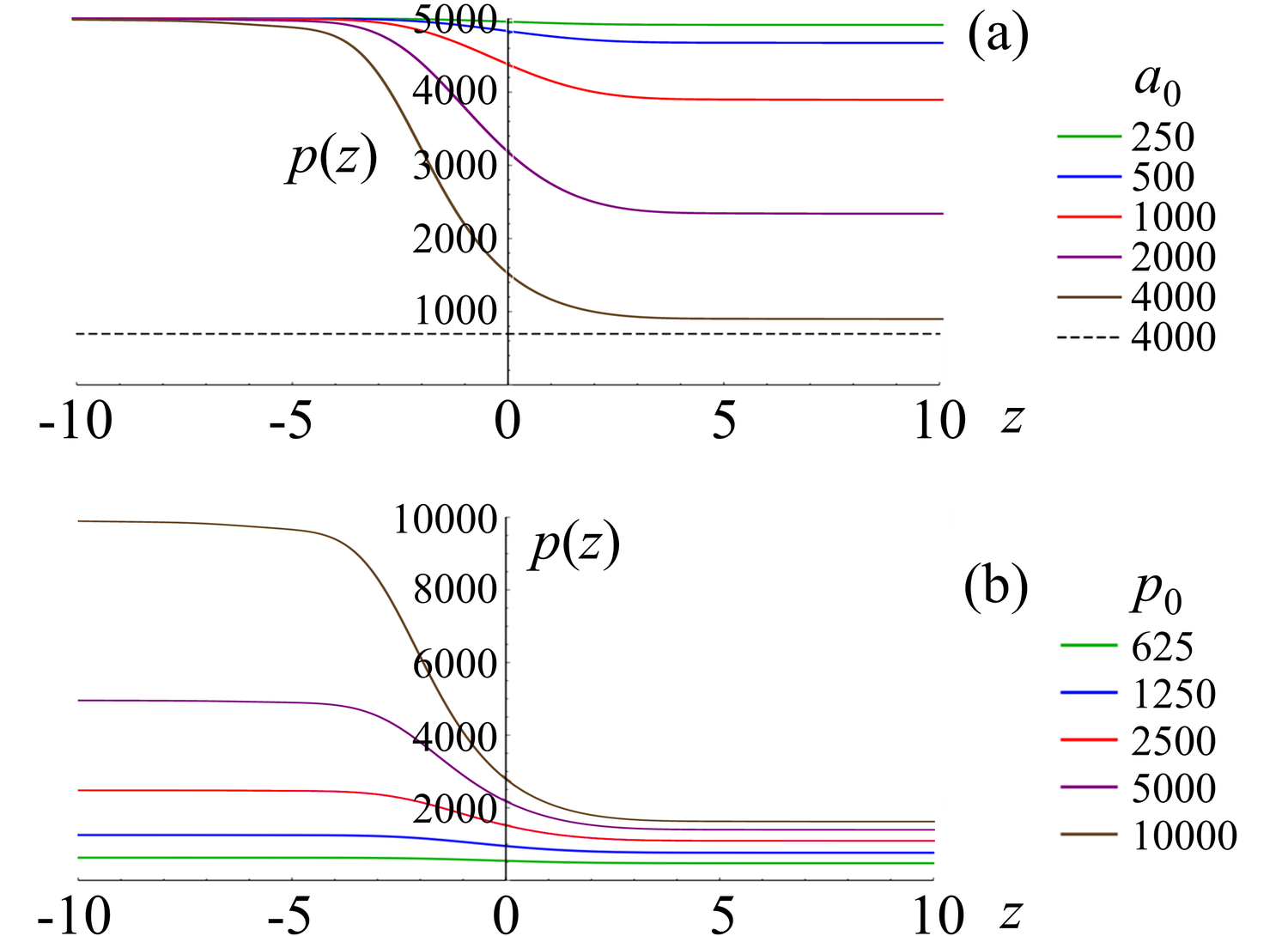}
\caption{%
a) Electron momentum $p(z)$ starting from $p_0=5000$
for different electromagnetic  field amplitudes $a_0$ varying from $250$ to $4000$.
Dashed line for Eq. (\ref{eq:gammaf}) at $a_0=4000$.
b) The same for $a_0=625$ and different $p_0$ varying from $625$ to $10000$.
The radiation friction parameter is $\varepsilon_{rad}=1.2\times 10^{-8}$;
the radial coordinate is $\rho_0=0.5$.
}
\label{fig:2}
\end{figure}

Fig.\,\ref{fig:2} shows the dependence of the particle momentum on the coordinate $z$ for different electromagnetic field amplitudes and initial momenta. 
In Panel (a) we plot the particle momentum versus the coordinate $z$ for the initial momentum equal to 5000 and for the electromagnetic  
field amplitude $a_0$ varying from $250$ to $4000$.
    The dashed line corresponds to the particle energy given by Eq. (\ref{eq:gammaf}) for $a_0=4000$. 
    Panel (b) presents dependence of the particle momentum on the coordinate $z$ for the electromagnetic field amplitude equal to $a_0=625 $ 
    and the initial momentum, $p_0$, varying from  $625$ to $10000$. The radiation friction parameter equals 
    $\varepsilon_{rad}=1.2\times 10^{-8}$; 
the radial coordinate is $\rho_0=0.5$. 
 As can be seen, for relatively low electromagnetic field amplitude the electron momentum after traversing the maximum field region  does not change significantly.

We have assumed above that the transverse component of the electron momentum is substantially smaller than the longitudinal component. 
The characteristic value of the transverse component of the momentum is approximately equal to the electromagnetic field amplitude $a_0$. 
It is easy to obtain that the transverse scattering of the electron becomes significant, if the electromagnetic 
field amplitude is large enough when $a_0 \gg  (\rho_0^2 \varepsilon_{rad})^{-1/3}$. 
To analyze the electron dynamics in the limit of 
strong electromagnetic field we present below the results of numerical integration 
of the equations of electron motion (\ref{eq:equmot-mom}).

\begin{figure}
    \includegraphics[width=0.66\columnwidth]{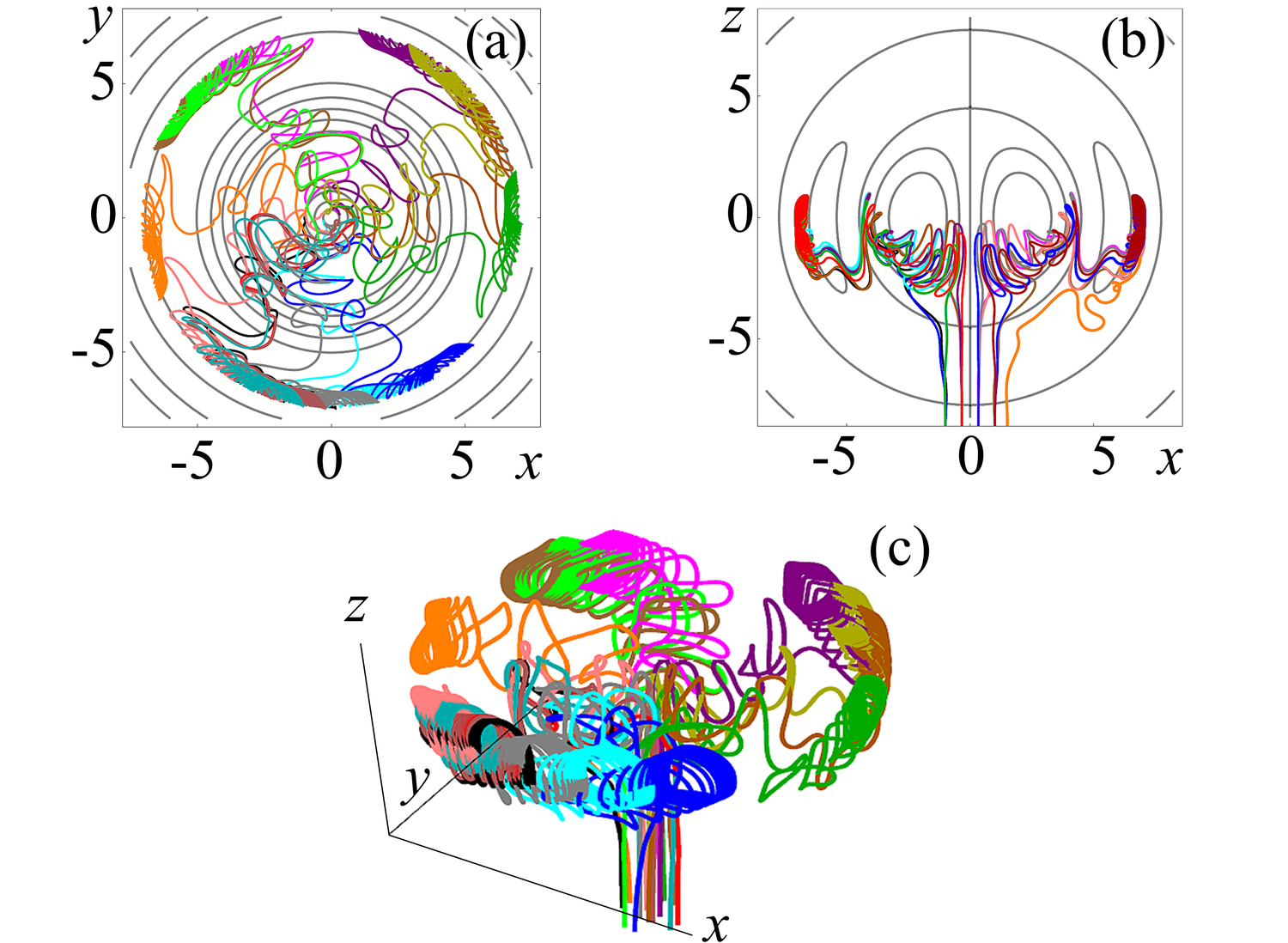}
\caption{%
Electron beam interaction with the TE-TM field configuration.
a) Electron trajectories (color) projected onto the $x,y$ plane.
Isocontours (black) of the magnetic field $B_{\phi}(\rho,z)|_{z=0}$.
b) The same projected onto the  $x, z$ plane.
Isocontours (black) of the electric field $B_{\phi}(\rho,z)|_{\rho=x}$.
c) The same in the cube $20^3$ centered at the origin.
The radiation friction parameter is $\varepsilon_{rad}=4\times 10^{-8}$;
the electromagnetic field amplitude is $a_m=\sqrt{8/9\pi}a_{TM}=3.1\times 10^3$.
}
\label{fig:3}
\end{figure}

Here we consider the interaction of a beam of ultrarelativistic electrons with the electromagnetic configuration of TE-TM polarization.
We note that in the limit of energy relatively low 
compared with the amplitude of the electromagnetic field, 
the electrons 
are reflected back by the ponderomotive force. In the high energy limit, when the electron energy substantially exceeds 
the ponderomotive potential and when the radiation friction effects are negligibly weak, the electrons propagate through 
the region of strong electromagnetic field being slightly scattered in the transverse direction. 
The situation drastically changes when the radiation friction force becomes dominant. This case is illustrated in Fig.  \ref{fig:3}.

Fig. \ref{fig:3} describes the electron beam interaction with the electromagnetic field of  the 
TE-TM
configuration. 
At $t=0$ the electron beam is mono-energetic with the
initial gamma factor equal to $\gamma_0=1000$, 
i.e. the initial electron energy is equal to 0.5 GeV.  
The radiation friction parameter is chosen to be $\varepsilon_{rad}=4\times 10^{-8}$. The electromagnetic field amplitude equals $a_m=20 \sqrt{8/9\pi}\varepsilon_{rad}^{-1/3}\approx 4\times 10^3$. 
In panel (a) the electron trajectories in the $x,y$ plane superimposed with the iso-contours 
of the magnetic field $B_{\phi}(\rho,z)|_{z=0}$ (or of the electric field $E_{\phi}(\rho,z)|_{z=0}$) are shown. 
Panel (b) shows the projection of the electron trajectories on the  $x, z$ plane superimposed with the contours of the constant value of the electric field $B_{\phi}(\rho,z)|_{\rho=x}$ (or of the electric field $E_{\phi}(\rho,z)|_{\rho=x}$) .
We see that all the electrons become trapped 
in the intervals of $-7.5<\rho<7.5$, $-3<z<1$.
The three dimensional pattern of the trajectories of the electron ensemble is presented in panel (c). 
   
\begin{figure}
    \includegraphics[width=0.66\columnwidth]{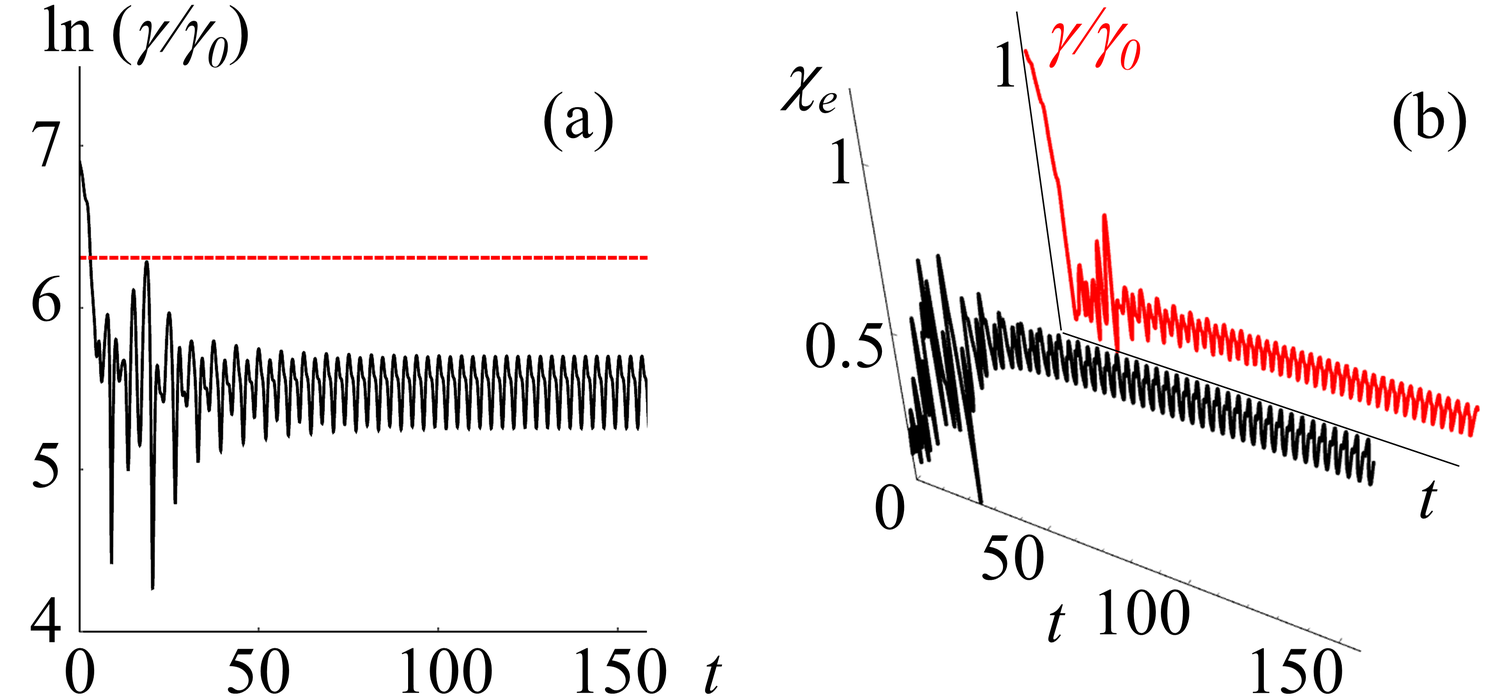}
\caption{%
a) Logarithm of the normalized electron energy, $\ln (\gamma/\gamma_0)$,  versus time.
Dashed line for  $(a_m/\varepsilon_{rad})^{1/4}$.
b) The parameter $\chi_e$  versus time compared with a).}
\label{fig:4}
\end{figure}

    In panel (a) of Fig. \ref{fig:4} we plot the dependence of  the logarithm of electron energy on time. 
    The dashed line corresponds 
    to the energy characterized by $\gamma_{rad}={\cal E}_{rad}/m_ec^2=(a_m/\varepsilon_{rad})^{1/4}$. 
    After a relatively short period of time, during which  the electron losses almost all its energy 
    and becomes trapped, the energy oscillates around the value below ${\cal E}_{rad}$. 
    This shows that the electron undergoes motion in the radiation dominated regime.
    In panel (b)  we show time dependencies of the normalized electron energy, $\ln (\gamma/\gamma_0)$,  and the parameter $\chi_e$  versus time.
When the electron loses its initial energy,
the parameter $\chi_e$ grows being less than unity, 
thus the classical electrodynamics approximation
assumed here 
is valid and one can neglect the photon recoil effects.
    
\begin{figure}
    \includegraphics[width=0.66\columnwidth]{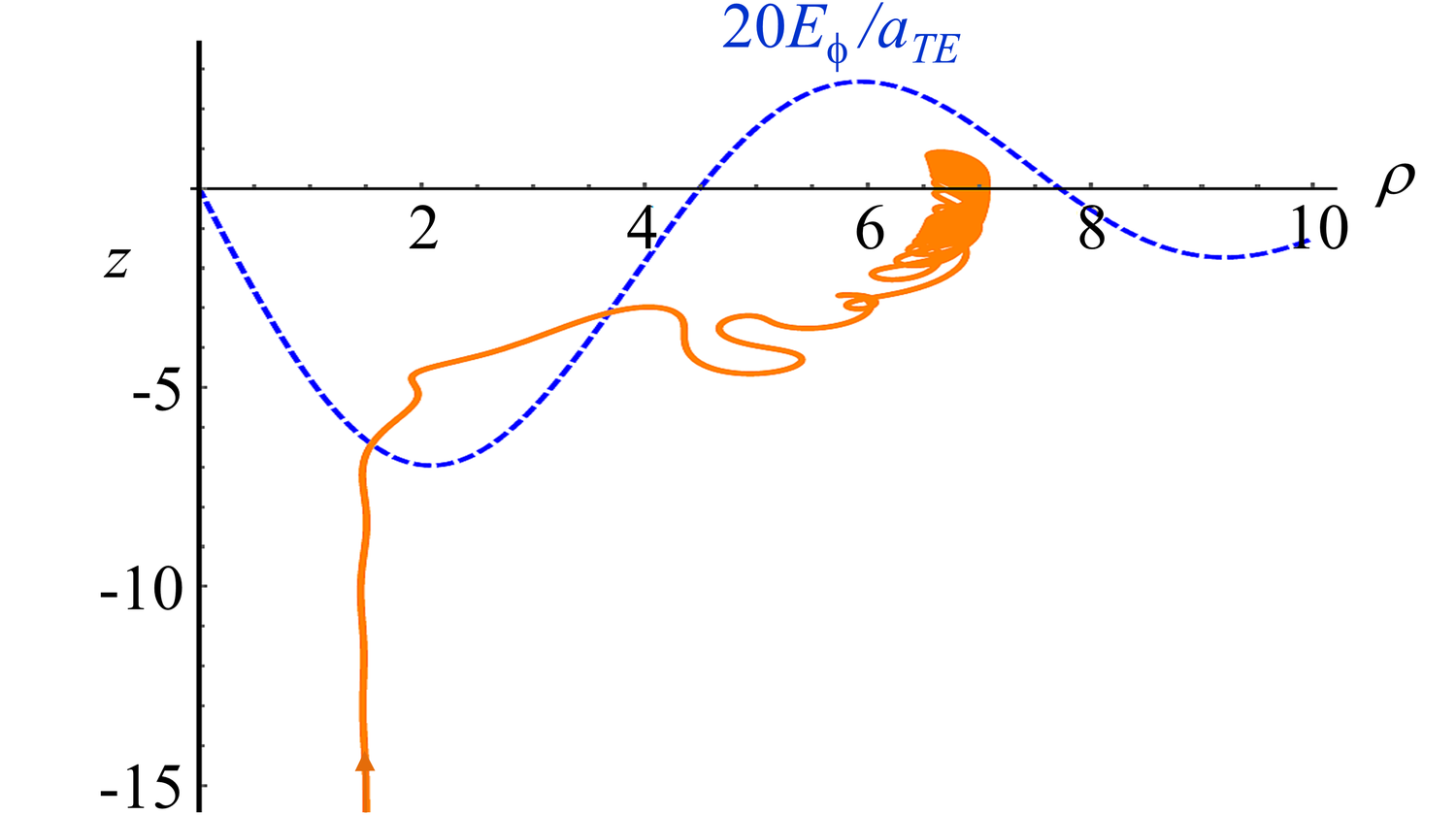}
    \caption{%
Trapped electron trajectory in the $\rho,z$ plane (solid, orange).
Dashed blue curve for the electric field $20 E_{\phi}(\rho,z)/a_{TE}$ at $z=0$.  }
\label{fig:5}
\end{figure}

    From Fig. \ref{fig:5}, where the trapped electron trajectory in the $\rho,z$ plane (orange, solid curve) 
    and the normalized component of the electric field 
$20 E_{\phi}(\rho,z=0)/a_{TE}$ versus the radial coordinate $ \rho$ (blue, dashed curve) 
    are plotted, it follows that the trapped electrons are located near the local 
maxima
of the electric field.
    Such radiating electron behavior happens in the
         scenario described earlier
\cite{Anom-bunch, Fedotov, Attr, Kirk}
within a low-dimensional
         geometry of electromagnetic configurations. 
As shown in Refs. \cite{Anom-bunch, Fedotov, Attr, Kirk, Survey, TZESVB},
electrons  can be captured for many laser periods due to
radiation friction impeding the ponderomotive force.
A collision of multiple ultra-intense electromagnetic waves
creates structurally determinate patterns 
in the electron phase space \cite{Survey, Vranic-2017, Gong-2017}
due to a counterplay of the ponderomotive force
and the friction-induced force.

From Fig. \ref{fig:3} it follows that, in the case when the radiation friction force becomes dominant, 
the TE-TM configuration becomes an efficient electromagnetic shield, 
which stops ultrarelativistic electrons over distances less than the electromagnetic mode wavelength. As a result, the electrons appear to be trapped with 
their trajectories located near  local maxima of the electric field.

The electromagnetic configurations 
considered above
can be formed by focusing the laser pulse with a parabolic mirror. 
Detailed theoretical calculations of this process can be found in Ref. \cite{Gonoskov-2012}. 
Within the Relativistic Flying Mirror (RFM) concept, 
formulated in Refs. \cite{RFM-2003, RFM-2013, RFM-2016}, 
the parabolic mirrors are formed as thin layers of relativistic electrons
in nonlinear wake waves excited in plasma behind an ultra short laser driver pulse. 
They have been proposed for reflection, focusing and intensification of another counter-propagating laser pulse. 
The advantage of the RFM can be seen in the 
denominator of Eq. (\ref{eq:radataxispINF}), where
$ p_0  \varepsilon_{rad}\, a_{0}^2 \rho_0^2$
can be rewritten in terms of the electromagnetic pulse power 
$\mathcal{P}$ and the pulse waist $w_0$ as 
$p_0 \rho_0^2 (32 r_e^2/3mc^3) (\lambdabar/w_0^2){\cal P}$. 
The laser power reflected from the RFM, ${\cal P}_r$, is nearly equal to $\cal{P}$ \cite{RFM-2016}.  
Since during the pulse reflection at the mirror moving with relativistic velocity both
the wavelength, $\lambdabar$, and spot size, $w_0$, are reduced, it can be seen that the net electron beam damping increases 
assuming the same values of $p_0$ and $\rho_0$. 
The use of the RFM 
enables achieving the extremely high amplitude electromagnetic field,
which is required for 
the interaction regimes considered 
here, in particular,
the efficient conversion of the relativistic electron energy to the energy of high energy photons.

We believe that the results obtained will further be used in designing the experiments for studying the extreme 
field limits in the interaction of lasers with ulrarelativistic electron 
beams \cite{design, Thomas_PRX-2012},
in developing high power ultra short gamma-ray sources \cite{Gamma_R-2012},
and in research on the probing of nonlinear quantum 
electrodynamics processes with high power lasers \cite{Vranic-2017, Gong-2017, Gonoskov_2016}.

SSB acknowledges support from the 
Office of Science of the US DOE under Contract No. DE-AC02-05CH11231.
JKK acknowledges support from JSPS KAKENHI Grant Number 16K05639.

\numberwithin{equation}{section}

\section*{Appendix}
\section{3D configuration of the electromagnetic field}
In a three-dimensional geometry, the electromagnetic field near the amplitude maximum,
depending on its polarization,
can be approximated either by the TM mode with 
Toroidal Magnetic and poloidal electric field or 
by the TE mode with 
Toroidal Electric and poloidal magnetic field or by the TM-TE mode made by the 
superposition of TM and TE modes. The toroidal magnetic and electric fields in spherical coordinates 
$R,\theta,\phi$ can be expressed via spherical harmonics \cite{Vainshtein-1988} as
\begin{equation}
\binom{{B_{\phi}}}{{E_{\phi}}}
=\frac{1}{\sqrt{R}} \binom{{a_{TM} \sin(t)}}{{a_{TE} \sin(t+\varphi_{TE})}}J_{n+1/2}(R)L_n^1(\cos(\theta)).
\label{eq:BE-phi-fields}
\end{equation}
Here and below we measure the electromagnetic field in the units of $m_e \omega c/e$, $a_{TM}$ and $a_{TE}$ 
are the normalized amplitudes of the TM and TE modes, and $\varphi_{TE}$ is 
the phase difference between them, 
$J_{\nu}(x)$ and $L_n^l(x)$ are, respectively, 
the Bessel function and associated Legendre polynomials \cite{AS}.
We assume azimuthal symmetry, i.e. $\partial_{\phi}=0$.
The variables $t$ and 
$R=\sqrt{x^2+y^2+z^2}$
are normalized by 
$\omega^{-1}$ and $k=c/\omega$, respectively.
In order to obtain the poloidal components,
one can use the relations between the Fourier components of electromagnetic fields.
The poloidal electric field in the TM mode is ${\bf E}=i k (\nabla \times {\bf B})$,
while
the poloidal magnetic field in the TE mode is  ${\bf B}=-i k (\nabla \times {\bf E})$.

In the highest symmetry nontrivial configuration with $n=1$ we have 
\begin{equation}
\binom{{B_{\phi}}}{{E_{\phi}}}
=
\sqrt{\frac{2}{\pi}}\binom{{a_{TM} \sin(t)}}{{a_{TE}} \sin(t+\varphi_{TE})}
\left (
\frac{\sin{R}-R \cos{R}}{R^2}
\right)\sin{\theta}.
\label{eq:BE-phi-low}
\end{equation}
It is convenient to write the expressions for toroidal 
components of the magnetic and electric field in cylindrical coordinates $(\rho, \phi, z)$,
$\rho=\sqrt{x^2+y^2}$:
\begin{equation}
\binom{{B_{\phi}}}{{E_{\phi}}}
=
\binom{{a_{TM} \sin(t)}}{{a_{TE}} \sin(t+\varphi_{TE})}
F_{\phi}(\rho,z),
\label{eq:BE-phi-cyl}
\end{equation}
where the function $F_{\phi}(\rho,z)$ determines the spatial distribution,
\begin{equation}
F_{\phi}(\rho,z)=\sqrt{\frac{2}{\pi}}\rho
 \left [
\frac{\sin{R}-R\cos{R}}{R^3}
\right],
\quad R=\sqrt{\rho^2+z^2}.
\label{eq:F-phi-cyl}
\end{equation}
The poloidal components of the magnetic and electric field
are given by the following expressions,
\begin{gather}
\binom{{E_{\rho}}}{{B_{\rho}}}
=
\binom{{a_{TM} \cos(t)}}{{a_{TE}} \cos(t+\varphi_{TE})}
F_{\rho}(\rho,z),
\label{eq:EB-rho-cyl}
\\
F_{\rho}(\rho,z)=\sqrt{\frac{2}{\pi}}z \, \rho\,\left[
\frac{3R \cos{R}+(R^2-3)\sin{R}}{R^{5}}
\right],
\label{eq:F-rho-cyl}
\\
\binom{{E_z}}{{B_{z}}}
=
\binom{{a_{TM} \cos(t)}}{{a_{TE}} \cos(t+\varphi_{TE})}
F_{z}(\rho,z),
\label{eq:EB-z-cyl}
\\
F_{z}(\rho,z)=
\sqrt{\frac{2}{\pi}}\rho \,\left[
\frac{(2 z^2-\rho^2)R \cos{R}-(2 z^2-\rho^2 + \rho^2 R^2)\sin{R}}{R^{5}}\right].
\label{eq:F-z-cyl}
\end{gather}

For the functions $F_{\phi}(\rho,z)$, $F_{\rho}(\rho,z)$, and $F_{z}(\rho,z)$, 
near the origin, in the limit of $\rho \to 0$ and $z \to 0$, we 
have 
\begin{gather}
F_{\phi}(\rho,z)=-\frac{1}{3}\sqrt{\frac{2}{\pi}}\rho+\frac{1}{15\sqrt{2\pi}}\rho z^2+\ldots\,,
\label{eq:Fphi}
%
\\
F_{\rho}(\rho,z)=-\frac{1}{15}\sqrt{\frac{2}{\pi}}\rho z+\ldots,
\label{eq:Frho}
%
\\
F_{z}(\rho,z)=-\frac{2}{3}\sqrt{\frac{2}{\pi}}\left(1-\frac{1}{5}\rho^2-\frac{1}{10}z^2\right)+\ldots\,.
\label{eq:Fz}
\end{gather}

\section{Electron Energy Losses due to the  Radiation Friction}

In the case of the electron moving parallel to  the $z$ axis,
if its trajectory is located at a small distance $\rho_0\ll 1$ from the $z$-axis, 
the electron motion is described by Eq. (5) above,
rewritten here as
\begin{gather}
\begin{split}
&dp/dz=-\varepsilon_{rad}\,p^2\,\Pi_z(\rho,z),
\\
&\Pi_z(\rho,z)=
\left(E_{\rho}(\rho,z)-B_{\phi}(\rho,z)\right)^2+
\left(E_{\phi}(\rho,z)+B_{\rho}(\rho,z)\right)^2.
\end{split}
\label{eq:radatZaxisBis}
\end{gather}
Here the function $\Pi_z(\rho,z)$ 
describes the space distribution of the radiation friction force. 
We assume that, for an ultrarelativistic particle moving parallel to the $z$ axis, $t=z$. 
Substituting Eqs. (\ref{eq:BE-phi-cyl}, \ref{eq:F-phi-cyl})
into Eq. (\ref{eq:radatZaxisBis}), we obtain
\begin{equation}
dp/dz=-\frac{2}{\pi}\varepsilon_{rad}\,p^2\,a_{0}^2 \rho_0^2\left\{\frac{z^2 [\sin (z)-z \cos (z)]^2+[(z^2-3) \sin (z)+3 z \cos (z)]^2}{z^8}\right\}.
\label{eq:radataxis}
\end{equation}

Integration of Eq. (\ref{eq:radataxis}) gives 
the dependence of the electron momentum on the coordinate $z$
\begin{equation}
p(z)=\frac{p_0}{1+\displaystyle (2/ \pi) p_0  \varepsilon_{rad}\, a_{0}^2 \rho_0^2 \Phi(z)},
\label{eq:radataxisp}
\end{equation}
where the function $\Phi(z)$ is given by
\begin{equation}
\Phi(z)=\frac{8  [\pi+ 2\, \text{Si}(2 z)]}{105}+
\frac{z (8 z^4 -12 z^2+270)\sin (2 z)+(16 z^6-8 z^4-186 z^2+135) \cos (2 z)-70 z^4-84 z^2-135}{210 z^7},
\label{eq:radataxisPhi}
\end{equation}
where $ {\rm Si}(x)=\int_0^x\sin(t) dt/t $ is the sine integral function \cite{AS}. 
The function $\Phi(z)$ changes from zero 
for $z\to -\infty$ to $16\pi/105$ for $z\to +\infty$. At $z=0$ the function  $\Phi(z)$ equals $8\pi/105$. In the vicinity of the point $z=0$ it 
linearly depends on the coordinate $z$:
\begin{equation}
\Phi(z)=\frac{8\pi}{105}+\frac{1}{9} z +\ldots,
\label{eq:radataxisPhiz0}
\end{equation}
i.e. its width is approximately equal to $72\pi/105\approx 2.15$.

Using expression for $\Pi_z$ and  Fig. (1) above, we can estimate the maximum value of the QED dimensionless parameter $\chi_e$. 
For the electron with the energy $\gamma$ moving along the $z$ direction in the TE-TM electromagnetic field $\chi_{e,{\rm max}}\sim 0.12\, \gamma a_0/a_S$.

\end{document}